\documentclass[aps, pra,floats,tightenlines, a4paper,twocolumn,10pt,longbibliography,superscriptaddress, nofootinbib]{revtex4-1}
\usepackage{bbm, amsmath, amssymb, amsthm, bm,textcomp, nicefrac,geometry,ragged2e}
\usepackage{graphicx,epstopdf,color,verbatim,enumitem}
\usepackage[dvipsnames]{xcolor}
\usepackage[bbgreekl]{mathbbol}
\usepackage{graphicx,epstopdf,color,verbatim,enumitem,ulem}
\definecolor{myurlcolor}{rgb}{0,0,0.7}
\definecolor{myrefcolor}{rgb}{0.8,0,0}
\usepackage[unicode=true,pdfusetitle, bookmarks=false,bookmarksnumbered=false,
bookmarksopen=false, breaklinks=false,pdfborder={0 0 0},backref=false,
colorlinks=true, linkcolor=myrefcolor,citecolor=myurlcolor,urlcolor=myurlcolor]{hyperref}
\usepackage{pbox,hyperref,array}

\newcommand{\ket}[1]{\left|#1\right\rangle}

\newcommand{\bra}[1]{\left\langle #1\right|}

\newcommand{\proj}[1]{\ket{#1}\bra{#1}}

\newcommand\cE{{\mathcal E}}

\newcommand{\be}{\begin{equation}}
\newcommand{\ee}{\end{equation}}
\newcommand{\bea}{\begin{eqnarray}}
\newcommand{\eea}{\end{eqnarray}}

\newcommand{\floor}[1]{\left\lfloor #1 \right\rfloor}

\begin{document}
\title{Optimal distributed sensing in noisy environments}
\author{P. Sekatski}
\thanks{These authors contributed equally}
\affiliation{Departement Physik, Universit\"at Basel, Klingelbergstra{\ss}e 82, 4056 Basel, Switzerland}
\author{S. W\"olk}
\thanks{These authors contributed equally}
 \affiliation{Institut f\"ur Theoretische Physik, Universit\"at Innsbruck, Technikerstra{\ss}e 21a, 6020 Innsbruck, Austria}
 \author{ W.~D\"ur}
  \affiliation{Institut f\"ur Theoretische Physik, Universit\"at Innsbruck, Technikerstra{\ss}e 21a, 6020 Innsbruck, Austria}\

\date{\today}

\begin{abstract}
We consider distributed sensing of non-local quantities. We introduce quantum enhanced protocols to directly measure any (scalar) field with a specific spatial dependence by placing sensors at appropriate positions and preparing a spatially distributed entangled quantum state. Our scheme has optimal Heisenberg scaling and is completely unaffected by noise on other processes with different spatial dependence than the signal. We consider both Fisher and Bayesian scenarios, and design states and settings to achieve optimal scaling. We explicitly demonstrate how to measure coefficients of spatial Taylor and Fourier series, and show that our approach can offer an exponential advantage as compared to strategies that do not make use of entanglement between different sites.
\end{abstract}
\maketitle

\paragraph*{Introduction.---}
High precision measurements of physical quantities are of fundamental importance in all branches of physics and beyond. Quantum metrology offers a quadratic scaling advantage over a classical approach, and has hence received tremendous attention in recent years. Most of the effort has concentrated on local estimation problems, where an unknown quantity such as field strength or frequency should be measured. Optimal schemes have been designed for different kinds of estimation problems, and demonstrated experimentally \cite{Giovannetti_2011,Toth_2014,Pezze_2018}.

In many physical problems, the quantity of interest is however not a local property, but has a characteristic spatial dependence such as e.g. the gradient (or higher moment) of a field, or a (spatial) Fourier coefficient. In this case, multiple measurements performed at different positions are required, i.e. one uses distributed sensors or sensor networks.
Such distributed sensors also allow one to increase resolution e.g. in classical imaging, where baseline telescopes are used. Recently quantum sensor networks have been introduced, and shown to offer an advantage in several problems: to measure field gradients \cite{Lanz_2013,Altenburg2017,Apellaniz2017}, to increase the accuracy of atomic clocks \cite{Komar_2014,Komar_2016}, or of interferometers and telescope networks \cite{Landini_2014,Ciampini_2016,Khabiboulline_2018,Khabiboulline_2018b} using entangled quantum states (see also \cite{Proctor_2017,Ge_2018,Eldredge_2018,Quian_2019,Humphreys_2013,Vidrighin_2014,Zhang_2014,Altenburg2016,Zhuang_2017,Kok_2017,Gessner_2018}). 
Current experimental capabilities (e.g. \cite{Blatt2008,Schindler_2013,childress_hanson_2013}) already allow the implementation of quantum sensor networks on the scale of a lab, and with
the emergence of quantum networks \cite{Kimble2008,Wehner_2018} large scale sensor networks shall become a promising application and a real possibility in the near future.
General quantum sensor networks are based on distributed multipartite entangled quantum states, and in addition to optimizing states, measurements and strategies also the positioning of sensors can be varied and optimized. Surprisingly, distributed entanglement between remote sensors does not necessarily help in the absence of noise and many repetitions (i.e. Fisher regime) when multiple quantities should be determined simultaneously \cite{Proctor_2017,Ragy_2016,Cyril2013,Altenburg_2018}. However, the practical applicability, in particular in the presence of noise and imperfections, is largely unexplored.

\begin{figure}[ht]
    \includegraphics[width=0.9 \columnwidth]{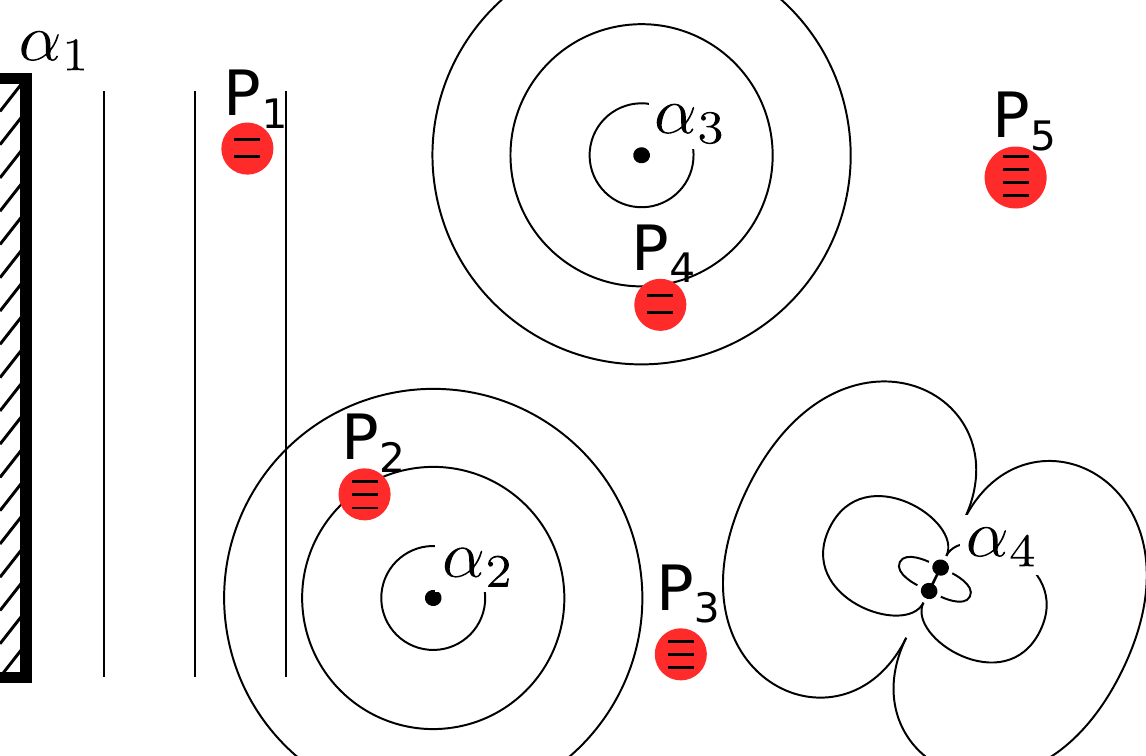}
    \caption{A sketch of a distributed sensing setup. Several sensors $P_\ell$ are located at different spatial positions. Different signals $\alpha_k$ have different spatial configurations of the field.}
    \label{figure_overview}
\end{figure}

Here we introduce quantum enhanced protocols to directly measure one or several scalar field components with a specific spatial dependence, e.g. of sources at specific positions, or coefficients of a spatial expansion function such as gradient or higher moment in a Taylor series (see Fig. \ref{figure_overview}). We explicitly determine positions and states in such a way that the states are capable of sensing only the components of interest, but are blind to all other processes with a different spatial dependence \footnote{More precisely, all spatial functions that are linearly independent of the signal. Typically $J+1$ sensor positions are required to be insensitive to $J$ signals.}. This offers very general and flexible schemes with multiple advantages: (i) the schemes have optimal Heisenberg scaling, i.e. they offer a quadratic scaling advantage over classical approaches; (ii) The schemes are completely insensitive to noise on other coefficients, and can offer up to an exponential improvement (in terms of number of locations) over strategies without distributed entanglement; (iii) The same (exponential) advantage can be maintained in noiseless single-shot experiments.

The applicability of a sensing scheme in a realistic environment, i.e. under the influence of noise and decoherence, is of particular importance. In local sensing it is known that noise severely limits the applicability of quantum strategies, and generically the quantum scaling advantage is reduced to a constant factor \cite{Fujiwara:08,Escher:11,Escher:12,Kolodynski:12,Sekatski2017quantummetrology,Sekatski2017PRX}. Only in some limited cases advanced techniques such as error correction or fast control allow one to maintain Heisenberg scaling \cite{Dur_2014,Arrad_2014,Kessler_2014,Sekatski_2016,Sekatski2017quantummetrology,Sekatski2017PRX,Zhou2018}. Our scheme is by construction blind to processes with a different spatial dependence, and hence completely insensitive to noise and fluctuations of these quantities. This is possible due to the additional freedom of placing sensors at arbitrary positions, and implies for example that gradients can be sensed despite arbitrarily large fluctuations of the global field. 
This offers a huge promise for near-term practical application of our methods.

We consider different kinds of estimation problems, including a frequency and phase estimation in both Fisher and Bayesian regime. We demonstrate our approach for specific examples, including spatial Taylor and Fourier series for sensing of magnetic fields, and also treat the case of various sources at different positions. In all cases we construct quantum states and positions with optimal scaling. In the Fisher regime the required measurements are in fact local, so one only needs to distribute entanglement for the preparation of the probe state.

\paragraph*{Estimating non-local expansion coefficients.---} Consider a network of $J$ sensors at positions ${\bf r}_j$
each consisting of $n_j$ qubits. The temporal evolution of each local system $j$ is determined by the local field $B({\bf r}_j)\equiv B_j$ and is given by
$U_j=\exp(i t B_j Z_j)$,
where $Z_j$ is the sum of Pauli-z operators for each qubit located at position ${\bf r}_j$, for convenience we will denote the state in the Z-basis by $\ket{+1}=\sigma_z\ket{+1}$ and $\ket{-1}=-\sigma_z\ket{-1}$. So that globally the evolution is given by $U=\bigotimes_j U_j$. The spatial configuration of the field can be expanded in some series of functions $\{{f}_k (r)\}$ as 
\be
 B({\bf r}) = \sum_k \alpha_k  f_k({\bf r}).
\ee
Typically, one is interested in the expansion, where different terms in the series $f_k({\bf r})$ corresponds to different physical processes or signals. Then the parameters $\alpha_k$ give the strength of each signal.

In terms of the parameters of interest $\alpha_k$, the temporal evolution can be rewritten as
$U= \exp\left(i t\sum_k \alpha_k G_k\right) = \exp\left(i\sum_k \Phi_k G_k\right)$ 
with the global phases $\Phi_k = t\alpha_k$ and the global generators
$G_k = \sum_j f_{kj}  Z_j$
where $f_{kj}=f_k(r_j)$, and we also define $\mathbf{f}_{k}= (f_{k1} \dots f_{kJ})$. In the Fisher regime, the optimal precision $(\Delta \Phi_{k_*})^2$ for determining a single parameter $\Phi_{k_*}$ is given by the Cramer-Rao bound
$(\Delta \Phi_{k_*})^2 \geq \frac{1}{4(\Delta G_{k_*})^2}$
where $(\Delta G_{k_0})^2$ is the variance of the generator on the probe state. When all the other parameters $\alpha_{k\neq k_*}$ are known, the optimal precision 
can be attained by preparing the probe in the superposition 
\be
\ket{\psi_{*}}= \left(\ket{\mathbf{s}_{*}} + \ket{-\mathbf{s}_{*}}  \right)/\sqrt{2}\label{eq:probe_state},
\ee
of the eigenstates of  $G_{k_*}$, that correspond to it's maximal and minimal eigenvalue. Because the generators are linear combinations of Pauli-z operators, all their eigenbasis are given by product states for all the qubits involved in the sensing. In particular, the extremal eigenstates are given by 
\be
\label{eq:optimal_eigenstate}
\ket{\pm \mathbf{ s}_{*}}=\bigotimes_{j=1}^J \ket{\textrm{sign}(\pm f_{k_* j})}^{\otimes n_j}.
\ee
The variance of $G_{k_*}$ for that state take the value
$(\Delta G_{k_*})^2 = \left(\sum_{j=1}^J n_j |f_{k_* j}|\right)^2$.


\paragraph*{Noise and decoherence free subspaces.---}

In the noiseless case, the presence of further components generated by  $G_{k}$ with $k\neq k_*$ is not detrimental to precision, as all the generators of the signals commute there exists a strategy that allows for optimal sensing of all the signals in the same time \cite{Proctor_2017}. However, this completely changes if some components $k\neq k_*$ are noisy.  If the value of the parameter $\alpha_k$ fluctuates, the coherence between any two eigenstates of $G_k$ with different eigenvalues is reduced during the evolution. In the worst case the coherence is completely washed out, as we shall see later. In order to be insensitive to this effect one can prepare the probe in a superposition $\ket{\psi_{k_*}}= \left(\ket{\mathbf{s}} + \ket{\mathbf{r}}  \right)/\sqrt{2}$  of two product states $\ket{\mathbf{s}} $ and $\ket{\mathbf{r}}$ that have the same eigenvalue for all the generators $G_k$ with $k\neq k_*$ but a different one for $G_{k_*}$. 

Let us now denote ${\bf s} = (s_1 ,\dots , s_J)$ with each $s_j$ being the eigenvalue $Z_j\ket{\bf s} = s_j \ket{\bf s}$ and ignore the degeneracy of this eigenvalue with respect to the permutations of the $n_j$ qubits. A priori $s_j$ and $r_j$ can take only integers in the interval $[-n_j,n_j]$. In many cases this is already sufficient as we show later. 
Non-integer values can be obtained by adding dynamical control, where at an intermediate time $t_j$ all spins at the corresponding site are switched, which results in an arbitrary effective value $s_j \in [-n_j,n_j]$. Effectively, the evolution for each sensor is slowed down. For time invariant parameters a single switching suffices, while a (fast) dynamical control allows one to do this in general. 

If the dynamical decoupling is conditional on an auxiliary qubit degree of freedom (e.g the permutation degeneracy we ignored), the slowing factor can be made different for the states $\ket{s_j}$ and $\ket{r_j}$ of the $j$-th sensor. Hence, we can take 
${\bf s}, {\bf r} \in O_{\bf n}=[-n_1, n_1]\times\dots\times [-n_J,n_J]$ 
to be any two vectors inside the $J$-orthotope (a box) $O_{\bf n}$.

The superposition state is then insensitive to the noises generated by $\alpha_k$ iff
\be\label{eq: noise insense}
{\bf f}_k^T({\bf s}-  {\bf r}) = 0 \quad \forall \, k\neq k_*,
\ee
and the quantum Fisher information of such a superposition with respect to the signal $\alpha_{k_*}$ is given by 
$4 \left(\Delta G_*\right)^2 = \left({\bf f}_{k_*}^T({\bf s} - {\bf r})\right)^2.$
Hence, to find the optimal strategy we need two vectors $\bf s$ and $\bf r$ that fulfill all the condition in Eq.\eqref{eq: noise insense} and give the maximal difference when projected on ${\bf f}_{k_*}$.

As we show in the appendix, the maximal sensitivity is attained by the state of Eq.\eqref{eq:probe_state} with
${\bf s}_* = \textrm{argmax}_{\bf s}\{{\bf f}_{k_*}^T {\bf s} |\,{\bf f}_{k}^T {\bf s} = 0 \quad \forall \quad k\neq k_*\}.$
Importantly, the optimal measurement strategy consists of performing local parity measurements and combining the classical measurement outcomes, and therefore does not require any entanglement.

Finally, let us also remark that if ${\bf f}_{k_*}$ is not linearly independent from $\{\bf f\}_{k\neq k_*}$, one can rearrange the positions of the sensors ${\bf r}_j$ or add new sensors to make the generators $G_*$ linearly independent from the others. The only case where this cannot be done is when the spacial dependence of the signal $f_{k_*}({\bf r})$ is linearly dependent from the noise processes $\{ f_{k}({\bf r})\}_{k\neq k_*}$. However in this case the signal is physically indistinguishable from the noise anyway. Similarly, if one wants to be independent from another noise process, one simply needs to add one additional sensor position. In general, $m+1$ sensors suffice to sense one signal with optimal Heisenberg-limit scaling, and be insensitive to $m$ independent noise process. We also note that the whole analysis remains true if the sensors are described by $d_j$-level systems and their evolution is governed by local Hamiltonians  $B({\bf r}_j) H_j$ with any nontrivial Hermitian operators $H_j$.

One can find also a whole subspace of states that is insensitive to a finite number of noise processes $G_k$, but sensitive to multiple signals. This is important in multiparameter  estimation problems, which can hence be simultaneously made insensitive against multiple noise processes. A simple strategy that already gives optimal Heisenberg scaling is to divide the sensing time among signals. A simultaneous sensing of different signals with an improved sensitivity is also possible. However, it is not clear how to devise optimal strategies, and if local measurement suffice in this case.


\paragraph*{Local estimation in noisy environments.---}

The scheme discussed above is universally applicable and offers optimal Heisenberg scaling, i.e. a quadratic improvement over classical strategies. By construction it is completely insensitive to noise on all coefficients ${\bf f}_k$. 
We now consider an explicit example to demonstrate that such a scheme that uses global entanglement can give an exponential improvement (in terms of number of sensors or noise functions) over schemes without entanglement between sensors, in contrast to the noiseless case \cite{Proctor_2017}.

\begin{figure}[ht]
    \includegraphics[width=0.8 \columnwidth]{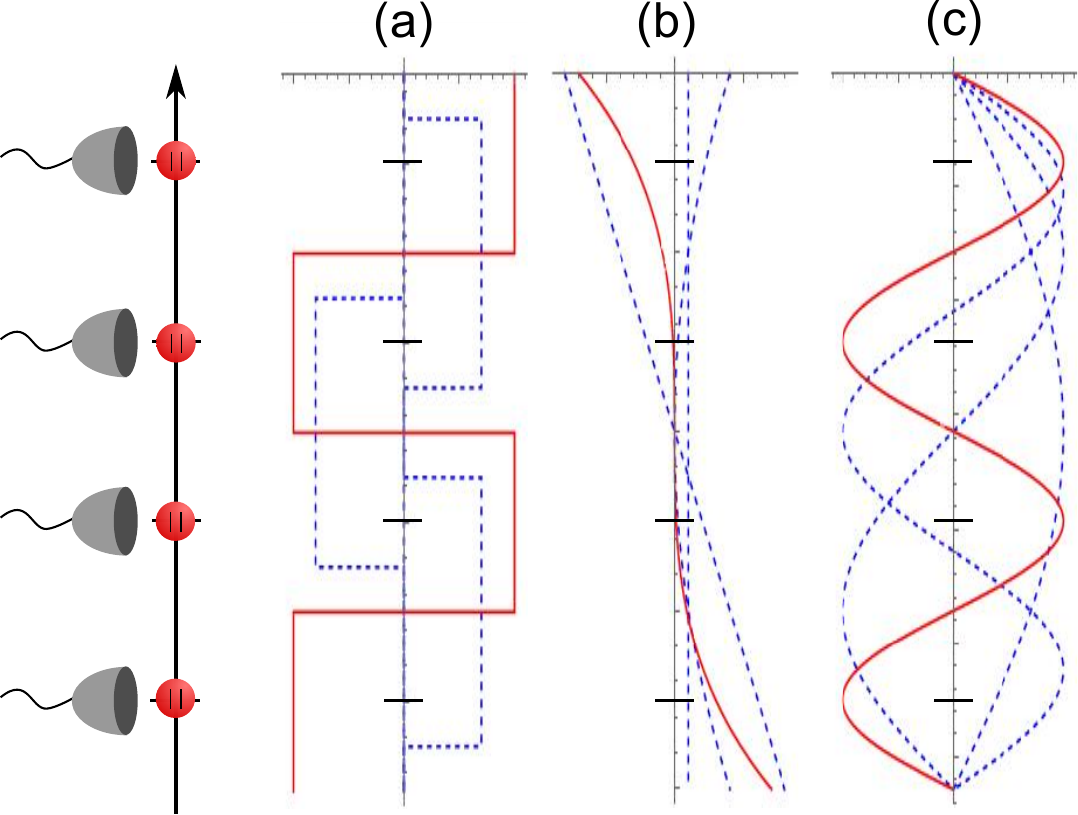}
    \caption{Examples of a few scenarios with sensors distributed on a line, discussed in the text. Noise is depicted in blue dashed and signal in red solid. (a) Short wavelength signal and local correlated noise. (b) Taylor series expansion (c) Fourier series expansion. }
    \label{Fig_examples}
\end{figure}

Consider $J$ equidistant sensors (with $J$ even) on a line at positions $r_j=j/J$, consisting of a single qubit each. 
The signal to sense is generated by the alternating function ${f}_{*}(r_j)=(-1)^j$, i.e. a high frequency Fourier coefficient where sensors are placed in the maximas.
In addition, we assume local correlated noise processes ${\bf f}_k$ with $f_k(r_j) = \delta_{k,j}+\delta_{k,j+1}$ acing on two neighbours only for $k=1,\dots,J-1$, see Fig. (\ref{Fig_examples}). Moreover, we consider the worst case scenario, where the effect noise on the probe state is captured by applying a twirling map, as described in appendix. 

The optimal state is then given by $(|{\bf s}_*\rangle + |-{\bf s}_*\rangle)/\sqrt{2}$ with $|{\bf s}_*\rangle=|1,-1,1, \ldots -1 \rangle$.  This state is already insensitive to all noise processes described above. In fact, $\ket{{\bf s}_*}$  and $\ket{-{\bf s}_*}$ form the only pair of states with matching eigenvalues for all noise generators $\{G_k\}_{k=1,\dots,J-1}$ -- all other states differ in at least one eigenvalue. This can be seen by noting that $G_{k\geq 1}$ force any pair of neighboring qubits to be anti-aligned $\ket{1}_k\ket{-1}_{k+1}$ or $\ket{-1}_k\ket{1}_{k+1}$ \footnote{We remark that this construction is also insensitive to global noise generated by a constant field ${\bf f}_0 = (1,\dots, 1)$.}.

Consider now a general state that is product w.r.t. different sensors, i.e. has no spatial entanglement. Such a state can be written as $\bigotimes_j (a_j|(-1)^j\rangle + b_j|(-1)^{j+1}\rangle)$ with $|a_k|^2+|b_k|^2=1$. The twirling map projects any such state onto the subspaces labeled by possible eigenvalues of the noise generators. However, we have just seen that the only nontrivial subspace is given by $\Pi_{\bf 0} = \proj{{\bf s}_*} + \proj{-{\bf s}_*}$, while all the other subspaces are of dimension one and cannot encode any information. It follows that a product state, can only sense the signal if it is projected onto $\Pi_{\bf 0}$, but the probability that it happens $P_{\bf 0} = \prod_{j=1}^J(a_j)^2 + \prod_{j=1}^J(b_j)^2\leq 2^{-(J-1)}$ decreases exponentially with $J$. Consequently, also the QFI for any strategy without entanglement between different sensors is $2^{-(J-1)}$ times lower than for the optimal state. This shows that in case of noise, spatial entanglement between sensors can provide up to an exponential advantage.


\paragraph*{Single event estimation (Bayesian).---}
The quantum Fisher information determines the precision of a metrology scheme in the "frequentist" scenario where the same experiment is repeated many time (aka Fisher regime) via the Cramer-Rao bound, and also in a single-shot scenario (aka Bayesian regime) if the parameter is approximately known and only a small deviation around this value needs to be determined \cite{Sekatski2017singlequbit}. 

In a general Bayesian scenario one starts with some initial knowledge of the parameters $\alpha_k$ described by the corresponding probability distributions $p_{\alpha_k}$ and aims to performs a {\it single} experimental run in such a way as to reduce the uncertainty about a certain parameter as much a possible. In some cases this procedure can then be repeated with the updated knowledge of the parameters, but in others one is limited to a single run. For example, the later can describe an event detection scenario, where all the information about an event has to be gathered in a short time window. We restrict our attention to these cases.

Generically, preparing the probes in a superposition of two eigenstates with the largest possible spectral gap $\Delta$ (maximal QFI), is not a good strategy for single-shot scenarios. The reason is simple, QFI only quantifies the rate at which the information about the signal can be gathered by the probes. But the total amount of information such binary superposition states encode is limited to one bit by the Holevo bound \cite{Holevo:73}. To overcome this limit eigenstates with intermediate eigenvalues become useful, see e.g. \cite{Berry2000,Sekatski2017singlequbit}. Hence in a generic Bayesian scenario both the QFI of the probe state and the number of different eigenvalues (and their actual values) are important. We will treat this in a separate publication \cite{WoelkSekatski_inpreparation}. 

In such a Bayesian event detection scenario, one may again have a {\it exponential} improvement with the number of sensors, however in this case even in a noiseless scenario. Indeed, the presence of additional signals, that do not need to be sensed but have very broad prior distributions, generate random unitary transformations of the probe's state and ultimately result in the same twirling map, as obtained in the noisy Fisher case above. 
The same example as outlined above can be used to demonstrate that there is an exponential loss factor $2^{-(J-1)}$ in the number of sensors $J$ for any scheme that uses only states that are not entangled among different sensors.


\paragraph*{Examples.---}
We now discuss several examples for using different expansion functions or spatial dependent signals, see Fig. (\ref{Fig_examples}). Additional details can be found in the appendix. 

As first example of generating functions in 1D, we consider the Taylor expansion with $f_k(r)=(r/r_0)^k$ where $r_0$ defines a length scale. $J$ sensors placed at arbitrary but mutually distinct locations $r_j$ are enough to discriminate the signals of $\lbrace{\bf f}_k\rbrace$ with  $0\leq k\leq J$. The determination of the optimal probe state Eq.\eqref{eq:probe_state} to determine $\Phi_{k^\ast}$ simplifies in the presence of $J-1$ noise sources. In this case we determine $\bf f_\perp$, with ${\bf f}_{k_\ast}=\bf f_\perp+ \bf f_\parallel$ and ${\bf f}_\perp^T {\bf f}_k=0$ $\forall k\neq k_\ast$, with the help of the Gram-Schmidt orthogonalization. The optimal state is then determined by  $\bf s \parallel \bf f_\perp$ since the decoherence free subspace is one dimensional. In the presence of noise with the same parity as the signal we find ${\bf f}_\perp\neq{\bf f}_{k_\ast}$ and thus the sensitivity is reduced although Heisenberg scaling is preserved. 

As second example we consider standing waves with boundary conditions $B(r)=0$ at positions $r/r_0=0$ and $r/r_0=1$. In this case, the generating functions are given by $f_k(r)=\sin[k\pi (r/r_0)]$. The Fourier coefficient $1\leq k\leq J$ can be uniquely determined by $J$ equidistant placed sensors (compare discrete Fourier transform). To determine the Fourier coefficient for $k_\ast=J$, it is optimal to place the sensors at the extremal points, that is $f_J(r_j)=\pm 1$, with alternating spins $s_j=(-1)^j n$ where we assumed $n$ qubits at each sensor $j$. This state is not only optimal in the noiseless case but also in the noisy case where all $f_k(r)$ with $k<J$ act as noise sources since ${\bf f}_{J}^T {\bf f}_k=0$.       
 
As a third example, consider $J$ field sources at different locations ${\bm r}_k$, where the distance dependence of the signal specifies the function $f_k$, $f_k({\bm r})=B_k g_k(|{\bm r}-{\bm r}_k|^{-\beta})$ with $\beta \geq 1$. The $f_k$ are linearly independent, and a choice of $J$ sensor locations provides a scheme with optimal Heisenberg scaling for one specific source that is insensitive to all $J-1$ other sources. If the degree of freedom of all noise sources is less than $J-1$ we obtain a multi-dimensional decoherence free subspace as outlined in the example in the appendix. In this case, the optimal $\bf s$ is not necessarily parallel to ${\bf f}_\perp$ but is given by some of the extremal points of the polytope $P_n$. Here, $P_n$ describes the decoherence free subspace bounded by the maximal number of qubits per sensor.



\paragraph*{Summary and conclusion.---}
In this work we have introduced general sensing schemes to directly measure a given quantity of interest with a certain spatial distribution. The schemes are designed for spatially distributed sensors in a quantum network, and allow one to e.g. measure the spatial dependence of the earth magnetic- or gravitational field on large scales. The methods are however equally well applicable within small sensing devices, where naturally different sensing systems are arranged in a specific spatial way, e.g. in a 1D line in a Paul trap in the case of trapped ions \cite{Blatt2008,Schindler_2013} or in a 3D arrangement as for NV centers in a crystal structure \cite{childress_hanson_2013}. In this case, the available control of such systems allows one to generate the required entangled states using established techniques, and to build magnetic field sensors for gradients or other spatial functions that are completely insensitive to spatially correlated noise processes. We have illustrated our methods for using standard expansion functions such as Taylor- or Fourier series where one of the expansion coefficients should be determined while others coefficients are effected by noise, but our methods can be adapted to provide insensitivity against dominant noise processes whenever they show a different spatial dependence than the signal to be sensed. This additional freedom gained by using spatially distributed entangled states is a powerful tool to maintain quantum enhancement even in the presence of noise and imperfections. As we have shown entangling the sensors within a network can offer up to an exponential advantage in precision.


\paragraph*{Acknowledgements.---} This work was supported by the Austrian Science Fund (FWF): P28000-N27, P30937-N27,  and SFB F40-FoQus F4012, by the Swiss National Science Foundation (SNSF) through Grant number PP00P2-150579, the Army Research Laboratory Center for Distributed Quantum Information via the project SciNet.

\bibliographystyle{apsrev4-1}
\bibliography{metrology}

\begin{thebibliography}{47}%
\makeatletter
\providecommand \@ifxundefined [1]{%
 \@ifx{#1\undefined}
}%
\providecommand \@ifnum [1]{%
 \ifnum #1\expandafter \@firstoftwo
 \else \expandafter \@secondoftwo
 \fi
}%
\providecommand \@ifx [1]{%
 \ifx #1\expandafter \@firstoftwo
 \else \expandafter \@secondoftwo
 \fi
}%
\providecommand \natexlab [1]{#1}%
\providecommand \enquote  [1]{``#1''}%
\providecommand \bibnamefont  [1]{#1}%
\providecommand \bibfnamefont [1]{#1}%
\providecommand \citenamefont [1]{#1}%
\providecommand \href@noop [0]{\@secondoftwo}%
\providecommand \href [0]{\begingroup \@sanitize@url \@href}%
\providecommand \@href[1]{\@@startlink{#1}\@@href}%
\providecommand \@@href[1]{\endgroup#1\@@endlink}%
\providecommand \@sanitize@url [0]{\catcode `\\12\catcode `\$12\catcode
  `\&12\catcode `\#12\catcode `\^12\catcode `\_12\catcode `\%12\relax}%
\providecommand \@@startlink[1]{}%
\providecommand \@@endlink[0]{}%
\providecommand \url  [0]{\begingroup\@sanitize@url \@url }%
\providecommand \@url [1]{\endgroup\@href {#1}{\urlprefix }}%
\providecommand \urlprefix  [0]{URL }%
\providecommand \Eprint [0]{\href }%
\providecommand \doibase [0]{http://dx.doi.org/}%
\providecommand \selectlanguage [0]{\@gobble}%
\providecommand \bibinfo  [0]{\@secondoftwo}%
\providecommand \bibfield  [0]{\@secondoftwo}%
\providecommand \translation [1]{[#1]}%
\providecommand \BibitemOpen [0]{}%
\providecommand \bibitemStop [0]{}%
\providecommand \bibitemNoStop [0]{.\EOS\space}%
\providecommand \EOS [0]{\spacefactor3000\relax}%
\providecommand \BibitemShut  [1]{\csname bibitem#1\endcsname}%
\let\auto@bib@innerbib\@empty
\bibitem [{\citenamefont {Giovannetti}\ \emph {et~al.}(2011)\citenamefont
  {Giovannetti}, \citenamefont {Lloyd},\ and\ \citenamefont
  {Maccone}}]{Giovannetti_2011}%
  \BibitemOpen
  \bibfield  {author} {\bibinfo {author} {\bibfnamefont {V.}~\bibnamefont
  {Giovannetti}}, \bibinfo {author} {\bibfnamefont {S.}~\bibnamefont {Lloyd}},
  \ and\ \bibinfo {author} {\bibfnamefont {L.}~\bibnamefont {Maccone}},\ }\href
  {https://doi.org/10.1038/nphoton.2011.35} {\bibfield  {journal} {\bibinfo
  {journal} {Nature Photonics}\ }\textbf {\bibinfo {volume} {5}},\ \bibinfo
  {pages} {222 EP } (\bibinfo {year} {2011})}\BibitemShut {NoStop}%
\bibitem [{\citenamefont {T{\'{o}}th}\ and\ \citenamefont
  {Apellaniz}(2014)}]{Toth_2014}%
  \BibitemOpen
  \bibfield  {author} {\bibinfo {author} {\bibfnamefont {G.}~\bibnamefont
  {T{\'{o}}th}}\ and\ \bibinfo {author} {\bibfnamefont {I.}~\bibnamefont
  {Apellaniz}},\ }\href {\doibase 10.1088/1751-8113/47/42/424006} {\bibfield
  {journal} {\bibinfo  {journal} {Journal of Physics A: Mathematical and
  Theoretical}\ }\textbf {\bibinfo {volume} {47}},\ \bibinfo {pages} {424006}
  (\bibinfo {year} {2014})}\BibitemShut {NoStop}%
\bibitem [{\citenamefont {Pezz\`e}\ \emph {et~al.}(2018)\citenamefont
  {Pezz\`e}, \citenamefont {Smerzi}, \citenamefont {Oberthaler}, \citenamefont
  {Schmied},\ and\ \citenamefont {Treutlein}}]{Pezze_2018}%
  \BibitemOpen
  \bibfield  {author} {\bibinfo {author} {\bibfnamefont {L.}~\bibnamefont
  {Pezz\`e}}, \bibinfo {author} {\bibfnamefont {A.}~\bibnamefont {Smerzi}},
  \bibinfo {author} {\bibfnamefont {M.~K.}\ \bibnamefont {Oberthaler}},
  \bibinfo {author} {\bibfnamefont {R.}~\bibnamefont {Schmied}}, \ and\
  \bibinfo {author} {\bibfnamefont {P.}~\bibnamefont {Treutlein}},\ }\href
  {\doibase 10.1103/RevModPhys.90.035005} {\bibfield  {journal} {\bibinfo
  {journal} {Rev. Mod. Phys.}\ }\textbf {\bibinfo {volume} {90}},\ \bibinfo
  {pages} {035005} (\bibinfo {year} {2018})}\BibitemShut {NoStop}%
\bibitem [{\citenamefont {Urizar-Lanz}\ \emph {et~al.}(2013)\citenamefont
  {Urizar-Lanz}, \citenamefont {Hyllus}, \citenamefont {Egusquiza},
  \citenamefont {Mitchell},\ and\ \citenamefont {T\'oth}}]{Lanz_2013}%
  \BibitemOpen
  \bibfield  {author} {\bibinfo {author} {\bibfnamefont {I.~n.}\ \bibnamefont
  {Urizar-Lanz}}, \bibinfo {author} {\bibfnamefont {P.}~\bibnamefont {Hyllus}},
  \bibinfo {author} {\bibfnamefont {I.~n.~L.}\ \bibnamefont {Egusquiza}},
  \bibinfo {author} {\bibfnamefont {M.~W.}\ \bibnamefont {Mitchell}}, \ and\
  \bibinfo {author} {\bibfnamefont {G.}~\bibnamefont {T\'oth}},\ }\href
  {\doibase 10.1103/PhysRevA.88.013626} {\bibfield  {journal} {\bibinfo
  {journal} {Phys. Rev. A}\ }\textbf {\bibinfo {volume} {88}},\ \bibinfo
  {pages} {013626} (\bibinfo {year} {2013})}\BibitemShut {NoStop}%
\bibitem [{\citenamefont {Altenburg}\ \emph {et~al.}(2017)\citenamefont
  {Altenburg}, \citenamefont {Oszmaniec}, \citenamefont {W\"olk},\ and\
  \citenamefont {G\"uhne}}]{Altenburg2017}%
  \BibitemOpen
  \bibfield  {author} {\bibinfo {author} {\bibfnamefont {S.}~\bibnamefont
  {Altenburg}}, \bibinfo {author} {\bibfnamefont {M.}~\bibnamefont
  {Oszmaniec}}, \bibinfo {author} {\bibfnamefont {S.}~\bibnamefont {W\"olk}}, \
  and\ \bibinfo {author} {\bibfnamefont {O.}~\bibnamefont {G\"uhne}},\ }\href
  {\doibase 10.1103/PhysRevA.96.042319} {\bibfield  {journal} {\bibinfo
  {journal} {Phys. Rev. A}\ }\textbf {\bibinfo {volume} {96}},\ \bibinfo
  {pages} {042319} (\bibinfo {year} {2017})}\BibitemShut {NoStop}%
\bibitem [{\citenamefont {Apellaniz}\ \emph {et~al.}(2018)\citenamefont
  {Apellaniz}, \citenamefont {Urizar-Lanz}, \citenamefont {Zimbor\'as},
  \citenamefont {Hyllus},\ and\ \citenamefont {T\'oth}}]{Apellaniz2017}%
  \BibitemOpen
  \bibfield  {author} {\bibinfo {author} {\bibfnamefont {I.}~\bibnamefont
  {Apellaniz}}, \bibinfo {author} {\bibfnamefont {I.~n.}\ \bibnamefont
  {Urizar-Lanz}}, \bibinfo {author} {\bibfnamefont {Z.}~\bibnamefont
  {Zimbor\'as}}, \bibinfo {author} {\bibfnamefont {P.}~\bibnamefont {Hyllus}},
  \ and\ \bibinfo {author} {\bibfnamefont {G.}~\bibnamefont {T\'oth}},\ }\href
  {\doibase 10.1103/PhysRevA.97.053603} {\bibfield  {journal} {\bibinfo
  {journal} {Phys. Rev. A}\ }\textbf {\bibinfo {volume} {97}},\ \bibinfo
  {pages} {053603} (\bibinfo {year} {2018})}\BibitemShut {NoStop}%
\bibitem [{\citenamefont {K{\'o}m{\'a}r}\ \emph {et~al.}(2014)\citenamefont
  {K{\'o}m{\'a}r}, \citenamefont {Kessler}, \citenamefont {Bishof},
  \citenamefont {Jiang}, \citenamefont {S{\o}rensen}, \citenamefont {Ye},\ and\
  \citenamefont {Lukin}}]{Komar_2014}%
  \BibitemOpen
  \bibfield  {author} {\bibinfo {author} {\bibfnamefont {P.}~\bibnamefont
  {K{\'o}m{\'a}r}}, \bibinfo {author} {\bibfnamefont {E.~M.}\ \bibnamefont
  {Kessler}}, \bibinfo {author} {\bibfnamefont {M.}~\bibnamefont {Bishof}},
  \bibinfo {author} {\bibfnamefont {L.}~\bibnamefont {Jiang}}, \bibinfo
  {author} {\bibfnamefont {A.~S.}\ \bibnamefont {S{\o}rensen}}, \bibinfo
  {author} {\bibfnamefont {J.}~\bibnamefont {Ye}}, \ and\ \bibinfo {author}
  {\bibfnamefont {M.~D.}\ \bibnamefont {Lukin}},\ }\href
  {https://doi.org/10.1038/nphys3000} {\bibfield  {journal} {\bibinfo
  {journal} {Nature Physics}\ }\textbf {\bibinfo {volume} {10}},\ \bibinfo
  {pages} {582 EP } (\bibinfo {year} {2014})},\ \bibinfo {note}
  {article}\BibitemShut {NoStop}%
\bibitem [{\citenamefont {K\'om\'ar}\ \emph {et~al.}(2016)\citenamefont
  {K\'om\'ar}, \citenamefont {Topcu}, \citenamefont {Kessler}, \citenamefont
  {Derevianko}, \citenamefont {Vuleti\ifmmode~\acute{c}\else \'{c}\fi{}},
  \citenamefont {Ye},\ and\ \citenamefont {Lukin}}]{Komar_2016}%
  \BibitemOpen
  \bibfield  {author} {\bibinfo {author} {\bibfnamefont {P.}~\bibnamefont
  {K\'om\'ar}}, \bibinfo {author} {\bibfnamefont {T.}~\bibnamefont {Topcu}},
  \bibinfo {author} {\bibfnamefont {E.~M.}\ \bibnamefont {Kessler}}, \bibinfo
  {author} {\bibfnamefont {A.}~\bibnamefont {Derevianko}}, \bibinfo {author}
  {\bibfnamefont {V.}~\bibnamefont {Vuleti\ifmmode~\acute{c}\else \'{c}\fi{}}},
  \bibinfo {author} {\bibfnamefont {J.}~\bibnamefont {Ye}}, \ and\ \bibinfo
  {author} {\bibfnamefont {M.~D.}\ \bibnamefont {Lukin}},\ }\href {\doibase
  10.1103/PhysRevLett.117.060506} {\bibfield  {journal} {\bibinfo  {journal}
  {Phys. Rev. Lett.}\ }\textbf {\bibinfo {volume} {117}},\ \bibinfo {pages}
  {060506} (\bibinfo {year} {2016})}\BibitemShut {NoStop}%
\bibitem [{\citenamefont {Landini}\ \emph {et~al.}(2014)\citenamefont
  {Landini}, \citenamefont {Fattori}, \citenamefont {Pezz{\`{e}}},\ and\
  \citenamefont {Smerzi}}]{Landini_2014}%
  \BibitemOpen
  \bibfield  {author} {\bibinfo {author} {\bibfnamefont {M.}~\bibnamefont
  {Landini}}, \bibinfo {author} {\bibfnamefont {M.}~\bibnamefont {Fattori}},
  \bibinfo {author} {\bibfnamefont {L.}~\bibnamefont {Pezz{\`{e}}}}, \ and\
  \bibinfo {author} {\bibfnamefont {A.}~\bibnamefont {Smerzi}},\ }\href
  {\doibase 10.1088/1367-2630/16/11/113074} {\bibfield  {journal} {\bibinfo
  {journal} {New Journal of Physics}\ }\textbf {\bibinfo {volume} {16}},\
  \bibinfo {pages} {113074} (\bibinfo {year} {2014})}\BibitemShut {NoStop}%
\bibitem [{\citenamefont {Ciampini}\ \emph {et~al.}(2016)\citenamefont
  {Ciampini}, \citenamefont {Spagnolo}, \citenamefont {Vitelli}, \citenamefont
  {Pezz{\`e}}, \citenamefont {Smerzi},\ and\ \citenamefont
  {Sciarrino}}]{Ciampini_2016}%
  \BibitemOpen
  \bibfield  {author} {\bibinfo {author} {\bibfnamefont {M.~A.}\ \bibnamefont
  {Ciampini}}, \bibinfo {author} {\bibfnamefont {N.}~\bibnamefont {Spagnolo}},
  \bibinfo {author} {\bibfnamefont {C.}~\bibnamefont {Vitelli}}, \bibinfo
  {author} {\bibfnamefont {L.}~\bibnamefont {Pezz{\`e}}}, \bibinfo {author}
  {\bibfnamefont {A.}~\bibnamefont {Smerzi}}, \ and\ \bibinfo {author}
  {\bibfnamefont {F.}~\bibnamefont {Sciarrino}},\ }\href
  {https://doi.org/10.1038/srep28881} {\bibfield  {journal} {\bibinfo
  {journal} {Scientific Reports}\ }\textbf {\bibinfo {volume} {6}},\ \bibinfo
  {pages} {28881 EP } (\bibinfo {year} {2016})},\ \bibinfo {note}
  {article}\BibitemShut {NoStop}%
\bibitem [{\citenamefont {Khabiboulline}\ \emph
  {et~al.}(2018{\natexlab{a}})\citenamefont {Khabiboulline}, \citenamefont
  {Borregaard}, \citenamefont {Greve},\ and\ \citenamefont
  {Lukin}}]{Khabiboulline_2018}%
  \BibitemOpen
  \bibfield  {author} {\bibinfo {author} {\bibfnamefont {E.~T.}\ \bibnamefont
  {Khabiboulline}}, \bibinfo {author} {\bibfnamefont {J.}~\bibnamefont
  {Borregaard}}, \bibinfo {author} {\bibfnamefont {K.~D.}\ \bibnamefont
  {Greve}}, \ and\ \bibinfo {author} {\bibfnamefont {M.~D.}\ \bibnamefont
  {Lukin}},\ }\href@noop {} {\enquote {\bibinfo {title} {Optical interferometry
  with quantum networks},}\ } (\bibinfo {year} {2018}{\natexlab{a}}),\ \Eprint
  {http://arxiv.org/abs/arXiv:1809.01659} {arXiv:1809.01659} \BibitemShut
  {NoStop}%
\bibitem [{\citenamefont {Khabiboulline}\ \emph
  {et~al.}(2018{\natexlab{b}})\citenamefont {Khabiboulline}, \citenamefont
  {Borregaard}, \citenamefont {Greve},\ and\ \citenamefont
  {Lukin}}]{Khabiboulline_2018b}%
  \BibitemOpen
  \bibfield  {author} {\bibinfo {author} {\bibfnamefont {E.~T.}\ \bibnamefont
  {Khabiboulline}}, \bibinfo {author} {\bibfnamefont {J.}~\bibnamefont
  {Borregaard}}, \bibinfo {author} {\bibfnamefont {K.~D.}\ \bibnamefont
  {Greve}}, \ and\ \bibinfo {author} {\bibfnamefont {M.~D.}\ \bibnamefont
  {Lukin}},\ }\href@noop {} {\enquote {\bibinfo {title} {Quantum-assisted
  telescope arrays},}\ } (\bibinfo {year} {2018}{\natexlab{b}}),\ \Eprint
  {http://arxiv.org/abs/arXiv:1809.03396} {arXiv:1809.03396} \BibitemShut
  {NoStop}%
\bibitem [{\citenamefont {Proctor}\ \emph {et~al.}(2018)\citenamefont
  {Proctor}, \citenamefont {Knott},\ and\ \citenamefont
  {Dunningham}}]{Proctor_2017}%
  \BibitemOpen
  \bibfield  {author} {\bibinfo {author} {\bibfnamefont {T.~J.}\ \bibnamefont
  {Proctor}}, \bibinfo {author} {\bibfnamefont {P.~A.}\ \bibnamefont {Knott}},
  \ and\ \bibinfo {author} {\bibfnamefont {J.~A.}\ \bibnamefont {Dunningham}},\
  }\href {\doibase 10.1103/PhysRevLett.120.080501} {\bibfield  {journal}
  {\bibinfo  {journal} {Phys. Rev. Lett.}\ }\textbf {\bibinfo {volume} {120}},\
  \bibinfo {pages} {080501} (\bibinfo {year} {2018})}\BibitemShut {NoStop}%
\bibitem [{\citenamefont {Ge}\ \emph {et~al.}(2018)\citenamefont {Ge},
  \citenamefont {Jacobs}, \citenamefont {Eldredge}, \citenamefont {Gorshkov},\
  and\ \citenamefont {Foss-Feig}}]{Ge_2018}%
  \BibitemOpen
  \bibfield  {author} {\bibinfo {author} {\bibfnamefont {W.}~\bibnamefont
  {Ge}}, \bibinfo {author} {\bibfnamefont {K.}~\bibnamefont {Jacobs}}, \bibinfo
  {author} {\bibfnamefont {Z.}~\bibnamefont {Eldredge}}, \bibinfo {author}
  {\bibfnamefont {A.~V.}\ \bibnamefont {Gorshkov}}, \ and\ \bibinfo {author}
  {\bibfnamefont {M.}~\bibnamefont {Foss-Feig}},\ }\href {\doibase
  10.1103/PhysRevLett.121.043604} {\bibfield  {journal} {\bibinfo  {journal}
  {Phys. Rev. Lett.}\ }\textbf {\bibinfo {volume} {121}},\ \bibinfo {pages}
  {043604} (\bibinfo {year} {2018})}\BibitemShut {NoStop}%
\bibitem [{\citenamefont {Eldredge}\ \emph {et~al.}(2018)\citenamefont
  {Eldredge}, \citenamefont {Foss-Feig}, \citenamefont {Gross}, \citenamefont
  {Rolston},\ and\ \citenamefont {Gorshkov}}]{Eldredge_2018}%
  \BibitemOpen
  \bibfield  {author} {\bibinfo {author} {\bibfnamefont {Z.}~\bibnamefont
  {Eldredge}}, \bibinfo {author} {\bibfnamefont {M.}~\bibnamefont {Foss-Feig}},
  \bibinfo {author} {\bibfnamefont {J.~A.}\ \bibnamefont {Gross}}, \bibinfo
  {author} {\bibfnamefont {S.~L.}\ \bibnamefont {Rolston}}, \ and\ \bibinfo
  {author} {\bibfnamefont {A.~V.}\ \bibnamefont {Gorshkov}},\ }\href {\doibase
  10.1103/PhysRevA.97.042337} {\bibfield  {journal} {\bibinfo  {journal} {Phys.
  Rev. A}\ }\textbf {\bibinfo {volume} {97}},\ \bibinfo {pages} {042337}
  (\bibinfo {year} {2018})}\BibitemShut {NoStop}%
\bibitem [{\citenamefont {Qian}\ \emph {et~al.}(2019)\citenamefont {Qian},
  \citenamefont {Eldredge}, \citenamefont {Ge}, \citenamefont {Pagano},
  \citenamefont {Monroe}, \citenamefont {Porto},\ and\ \citenamefont
  {Gorshkov}}]{Quian_2019}%
  \BibitemOpen
  \bibfield  {author} {\bibinfo {author} {\bibfnamefont {K.}~\bibnamefont
  {Qian}}, \bibinfo {author} {\bibfnamefont {Z.}~\bibnamefont {Eldredge}},
  \bibinfo {author} {\bibfnamefont {W.}~\bibnamefont {Ge}}, \bibinfo {author}
  {\bibfnamefont {G.}~\bibnamefont {Pagano}}, \bibinfo {author} {\bibfnamefont
  {C.}~\bibnamefont {Monroe}}, \bibinfo {author} {\bibfnamefont {J.~V.}\
  \bibnamefont {Porto}}, \ and\ \bibinfo {author} {\bibfnamefont {A.~V.}\
  \bibnamefont {Gorshkov}},\ }\href@noop {} {\enquote {\bibinfo {title}
  {Heisenberg-scaling measurement protocol for analytic functions with quantum
  sensor networks},}\ } (\bibinfo {year} {2019}),\ \Eprint
  {http://arxiv.org/abs/arXiv:1901.09042} {arXiv:1901.09042} \BibitemShut
  {NoStop}%
\bibitem [{\citenamefont {Humphreys}\ \emph {et~al.}(2013)\citenamefont
  {Humphreys}, \citenamefont {Barbieri}, \citenamefont {Datta},\ and\
  \citenamefont {Walmsley}}]{Humphreys_2013}%
  \BibitemOpen
  \bibfield  {author} {\bibinfo {author} {\bibfnamefont {P.~C.}\ \bibnamefont
  {Humphreys}}, \bibinfo {author} {\bibfnamefont {M.}~\bibnamefont {Barbieri}},
  \bibinfo {author} {\bibfnamefont {A.}~\bibnamefont {Datta}}, \ and\ \bibinfo
  {author} {\bibfnamefont {I.~A.}\ \bibnamefont {Walmsley}},\ }\href {\doibase
  10.1103/PhysRevLett.111.070403} {\bibfield  {journal} {\bibinfo  {journal}
  {Phys. Rev. Lett.}\ }\textbf {\bibinfo {volume} {111}},\ \bibinfo {pages}
  {070403} (\bibinfo {year} {2013})}\BibitemShut {NoStop}%
\bibitem [{\citenamefont {Vidrighin}\ \emph {et~al.}(2014)\citenamefont
  {Vidrighin}, \citenamefont {Donati}, \citenamefont {Genoni}, \citenamefont
  {Jin}, \citenamefont {Kolthammer}, \citenamefont {Kim}, \citenamefont
  {Datta}, \citenamefont {Barbieri},\ and\ \citenamefont
  {Walmsley}}]{Vidrighin_2014}%
  \BibitemOpen
  \bibfield  {author} {\bibinfo {author} {\bibfnamefont {M.~D.}\ \bibnamefont
  {Vidrighin}}, \bibinfo {author} {\bibfnamefont {G.}~\bibnamefont {Donati}},
  \bibinfo {author} {\bibfnamefont {M.~G.}\ \bibnamefont {Genoni}}, \bibinfo
  {author} {\bibfnamefont {X.-M.}\ \bibnamefont {Jin}}, \bibinfo {author}
  {\bibfnamefont {W.~S.}\ \bibnamefont {Kolthammer}}, \bibinfo {author}
  {\bibfnamefont {M.~S.}\ \bibnamefont {Kim}}, \bibinfo {author} {\bibfnamefont
  {A.}~\bibnamefont {Datta}}, \bibinfo {author} {\bibfnamefont
  {M.}~\bibnamefont {Barbieri}}, \ and\ \bibinfo {author} {\bibfnamefont
  {I.~A.}\ \bibnamefont {Walmsley}},\ }\href
  {https://doi.org/10.1038/ncomms4532} {\bibfield  {journal} {\bibinfo
  {journal} {Nature Communications}\ }\textbf {\bibinfo {volume} {5}},\
  \bibinfo {pages} {3532 EP } (\bibinfo {year} {2014})},\ \bibinfo {note}
  {article}\BibitemShut {NoStop}%
\bibitem [{\citenamefont {Zhang}\ and\ \citenamefont {Fan}(2014)}]{Zhang_2014}%
  \BibitemOpen
  \bibfield  {author} {\bibinfo {author} {\bibfnamefont {Y.-R.}\ \bibnamefont
  {Zhang}}\ and\ \bibinfo {author} {\bibfnamefont {H.}~\bibnamefont {Fan}},\
  }\href {\doibase 10.1103/PhysRevA.90.043818} {\bibfield  {journal} {\bibinfo
  {journal} {Phys. Rev. A}\ }\textbf {\bibinfo {volume} {90}},\ \bibinfo
  {pages} {043818} (\bibinfo {year} {2014})}\BibitemShut {NoStop}%
\bibitem [{\citenamefont {Altenburg}\ \emph {et~al.}(2016)\citenamefont
  {Altenburg}, \citenamefont {W\"olk}, \citenamefont {T\'oth},\ and\
  \citenamefont {G\"uhne}}]{Altenburg2016}%
  \BibitemOpen
  \bibfield  {author} {\bibinfo {author} {\bibfnamefont {S.}~\bibnamefont
  {Altenburg}}, \bibinfo {author} {\bibfnamefont {S.}~\bibnamefont {W\"olk}},
  \bibinfo {author} {\bibfnamefont {G.}~\bibnamefont {T\'oth}}, \ and\ \bibinfo
  {author} {\bibfnamefont {O.}~\bibnamefont {G\"uhne}},\ }\href {\doibase
  10.1103/PhysRevA.94.052306} {\bibfield  {journal} {\bibinfo  {journal} {Phys.
  Rev. A}\ }\textbf {\bibinfo {volume} {94}},\ \bibinfo {pages} {052306}
  (\bibinfo {year} {2016})}\BibitemShut {NoStop}%
\bibitem [{\citenamefont {Zhuang}\ \emph {et~al.}(2017)\citenamefont {Zhuang},
  \citenamefont {Zhang},\ and\ \citenamefont {Shapiro}}]{Zhuang_2017}%
  \BibitemOpen
  \bibfield  {author} {\bibinfo {author} {\bibfnamefont {Q.}~\bibnamefont
  {Zhuang}}, \bibinfo {author} {\bibfnamefont {Z.}~\bibnamefont {Zhang}}, \
  and\ \bibinfo {author} {\bibfnamefont {J.}~\bibnamefont {Shapiro}},\ }\href
  {\doibase 10.1103/PhysRevA.97.032329} {\bibfield  {journal} {\bibinfo
  {journal} {Physical Review A}\ }\textbf {\bibinfo {volume} {97}} (\bibinfo
  {year} {2017}),\ 10.1103/PhysRevA.97.032329}\BibitemShut {NoStop}%
\bibitem [{\citenamefont {Kok}\ \emph {et~al.}(2017)\citenamefont {Kok},
  \citenamefont {Dunningham},\ and\ \citenamefont {Ralph}}]{Kok_2017}%
  \BibitemOpen
  \bibfield  {author} {\bibinfo {author} {\bibfnamefont {P.}~\bibnamefont
  {Kok}}, \bibinfo {author} {\bibfnamefont {J.}~\bibnamefont {Dunningham}}, \
  and\ \bibinfo {author} {\bibfnamefont {J.~F.}\ \bibnamefont {Ralph}},\ }\href
  {\doibase 10.1103/PhysRevA.95.012326} {\bibfield  {journal} {\bibinfo
  {journal} {Phys. Rev. A}\ }\textbf {\bibinfo {volume} {95}},\ \bibinfo
  {pages} {012326} (\bibinfo {year} {2017})}\BibitemShut {NoStop}%
\bibitem [{\citenamefont {Gessner}\ \emph {et~al.}(2018)\citenamefont
  {Gessner}, \citenamefont {Pezz\`e},\ and\ \citenamefont
  {Smerzi}}]{Gessner_2018}%
  \BibitemOpen
  \bibfield  {author} {\bibinfo {author} {\bibfnamefont {M.}~\bibnamefont
  {Gessner}}, \bibinfo {author} {\bibfnamefont {L.}~\bibnamefont {Pezz\`e}}, \
  and\ \bibinfo {author} {\bibfnamefont {A.}~\bibnamefont {Smerzi}},\ }\href
  {\doibase 10.1103/PhysRevLett.121.130503} {\bibfield  {journal} {\bibinfo
  {journal} {Phys. Rev. Lett.}\ }\textbf {\bibinfo {volume} {121}},\ \bibinfo
  {pages} {130503} (\bibinfo {year} {2018})}\BibitemShut {NoStop}%
\bibitem [{\citenamefont {Blatt}\ and\ \citenamefont
  {Wineland}(2008)}]{Blatt2008}%
  \BibitemOpen
  \bibfield  {author} {\bibinfo {author} {\bibfnamefont {R.}~\bibnamefont
  {Blatt}}\ and\ \bibinfo {author} {\bibfnamefont {D.}~\bibnamefont
  {Wineland}},\ }\href {https://doi.org/10.1038/nature07125} {\bibfield
  {journal} {\bibinfo  {journal} {Nature}\ }\textbf {\bibinfo {volume} {453}},\
  \bibinfo {pages} {1008 EP } (\bibinfo {year} {2008})}\BibitemShut {NoStop}%
\bibitem [{\citenamefont {Schindler}\ \emph {et~al.}(2013)\citenamefont
  {Schindler}, \citenamefont {Nigg}, \citenamefont {Monz}, \citenamefont
  {Barreiro}, \citenamefont {Martinez}, \citenamefont {Wang}, \citenamefont
  {Quint}, \citenamefont {Brandl}, \citenamefont {Nebendahl}, \citenamefont
  {Roos}, \citenamefont {Chwalla}, \citenamefont {Hennrich},\ and\
  \citenamefont {Blatt}}]{Schindler_2013}%
  \BibitemOpen
  \bibfield  {author} {\bibinfo {author} {\bibfnamefont {P.}~\bibnamefont
  {Schindler}}, \bibinfo {author} {\bibfnamefont {D.}~\bibnamefont {Nigg}},
  \bibinfo {author} {\bibfnamefont {T.}~\bibnamefont {Monz}}, \bibinfo {author}
  {\bibfnamefont {J.~T.}\ \bibnamefont {Barreiro}}, \bibinfo {author}
  {\bibfnamefont {E.}~\bibnamefont {Martinez}}, \bibinfo {author}
  {\bibfnamefont {S.~X.}\ \bibnamefont {Wang}}, \bibinfo {author}
  {\bibfnamefont {S.}~\bibnamefont {Quint}}, \bibinfo {author} {\bibfnamefont
  {M.~F.}\ \bibnamefont {Brandl}}, \bibinfo {author} {\bibfnamefont
  {V.}~\bibnamefont {Nebendahl}}, \bibinfo {author} {\bibfnamefont {C.~F.}\
  \bibnamefont {Roos}}, \bibinfo {author} {\bibfnamefont {M.}~\bibnamefont
  {Chwalla}}, \bibinfo {author} {\bibfnamefont {M.}~\bibnamefont {Hennrich}}, \
  and\ \bibinfo {author} {\bibfnamefont {R.}~\bibnamefont {Blatt}},\ }\href
  {\doibase 10.1088/1367-2630/15/12/123012} {\bibfield  {journal} {\bibinfo
  {journal} {New Journal of Physics}\ }\textbf {\bibinfo {volume} {15}},\
  \bibinfo {pages} {123012} (\bibinfo {year} {2013})}\BibitemShut {NoStop}%
\bibitem [{\citenamefont {Childress}\ and\ \citenamefont
  {Hanson}(2013)}]{childress_hanson_2013}%
  \BibitemOpen
  \bibfield  {author} {\bibinfo {author} {\bibfnamefont {L.}~\bibnamefont
  {Childress}}\ and\ \bibinfo {author} {\bibfnamefont {R.}~\bibnamefont
  {Hanson}},\ }\href {\doibase 10.1557/mrs.2013.20} {\bibfield  {journal}
  {\bibinfo  {journal} {MRS Bulletin}\ }\textbf {\bibinfo {volume} {38}},\
  \bibinfo {pages} {134–138} (\bibinfo {year} {2013})}\BibitemShut {NoStop}%
\bibitem [{\citenamefont {Kimble}(2008)}]{Kimble2008}%
  \BibitemOpen
  \bibfield  {author} {\bibinfo {author} {\bibfnamefont {H.~J.}\ \bibnamefont
  {Kimble}},\ }\href {https://doi.org/10.1038/nature07127} {\bibfield
  {journal} {\bibinfo  {journal} {Nature}\ }\textbf {\bibinfo {volume} {453}},\
  \bibinfo {pages} {1023 EP } (\bibinfo {year} {2008})}\BibitemShut {NoStop}%
\bibitem [{\citenamefont {Wehner}\ \emph {et~al.}(2018)\citenamefont {Wehner},
  \citenamefont {Elkouss},\ and\ \citenamefont {Hanson}}]{Wehner_2018}%
  \BibitemOpen
  \bibfield  {author} {\bibinfo {author} {\bibfnamefont {S.}~\bibnamefont
  {Wehner}}, \bibinfo {author} {\bibfnamefont {D.}~\bibnamefont {Elkouss}}, \
  and\ \bibinfo {author} {\bibfnamefont {R.}~\bibnamefont {Hanson}},\ }\href
  {\doibase 10.1126/science.aam9288} {\bibfield  {journal} {\bibinfo  {journal}
  {Science}\ }\textbf {\bibinfo {volume} {362}} (\bibinfo {year} {2018}),\
  10.1126/science.aam9288}\BibitemShut {NoStop}%
\bibitem [{\citenamefont {Ragy}\ \emph {et~al.}(2016)\citenamefont {Ragy},
  \citenamefont {Jarzyna},\ and\ \citenamefont
  {Demkowicz-Dobrza\ifmmode~\acute{n}\else \'{n}\fi{}ski}}]{Ragy_2016}%
  \BibitemOpen
  \bibfield  {author} {\bibinfo {author} {\bibfnamefont {S.}~\bibnamefont
  {Ragy}}, \bibinfo {author} {\bibfnamefont {M.}~\bibnamefont {Jarzyna}}, \
  and\ \bibinfo {author} {\bibfnamefont {R.}~\bibnamefont
  {Demkowicz-Dobrza\ifmmode~\acute{n}\else \'{n}\fi{}ski}},\ }\href {\doibase
  10.1103/PhysRevA.94.052108} {\bibfield  {journal} {\bibinfo  {journal} {Phys.
  Rev. A}\ }\textbf {\bibinfo {volume} {94}},\ \bibinfo {pages} {052108}
  (\bibinfo {year} {2016})}\BibitemShut {NoStop}%
\bibitem [{\citenamefont {Cyril}\ \emph {et~al.}(2013)\citenamefont {Cyril},
  \citenamefont {Tommaso},\ and\ \citenamefont {G.}}]{Cyril2013}%
  \BibitemOpen
  \bibfield  {author} {\bibinfo {author} {\bibfnamefont {V.}~\bibnamefont
  {Cyril}}, \bibinfo {author} {\bibfnamefont {T.}~\bibnamefont {Tommaso}}, \
  and\ \bibinfo {author} {\bibfnamefont {G.~M.}\ \bibnamefont {G.}},\ }\enquote
  {\bibinfo {title} {qmetro},}\ \ (\bibinfo {year} {2013})\ Chap.\ \bibinfo
  {chapter} {Quantum estimation of a two-phase spin rotation}, p.~\bibinfo
  {pages} {12}\BibitemShut {NoStop}%
\bibitem [{\citenamefont {Altenburg}\ and\ \citenamefont
  {W\"olk}(2018)}]{Altenburg_2018}%
  \BibitemOpen
  \bibfield  {author} {\bibinfo {author} {\bibfnamefont {S.}~\bibnamefont
  {Altenburg}}\ and\ \bibinfo {author} {\bibfnamefont {S.}~\bibnamefont
  {W\"olk}},\ }\href {\doibase 10.1088/1402-4896/aaeca1} {\bibfield  {journal}
  {\bibinfo  {journal} {Physica Scripta}\ }\textbf {\bibinfo {volume} {94}},\
  \bibinfo {pages} {014001} (\bibinfo {year} {2018})}\BibitemShut {NoStop}%
\bibitem [{\citenamefont {Fujiwara}\ and\ \citenamefont
  {Imai}(2008)}]{Fujiwara:08}%
  \BibitemOpen
  \bibfield  {author} {\bibinfo {author} {\bibfnamefont {A.}~\bibnamefont
  {Fujiwara}}\ and\ \bibinfo {author} {\bibfnamefont {H.}~\bibnamefont
  {Imai}},\ }\href {\doibase 10.1088/1751-8113/41/25/255304} {\bibfield
  {journal} {\bibinfo  {journal} {Journal of Physics A: Mathematical and
  Theoretical}\ }\textbf {\bibinfo {volume} {41}},\ \bibinfo {pages} {255304}
  (\bibinfo {year} {2008})}\BibitemShut {NoStop}%
\bibitem [{\citenamefont {Escher}\ \emph {et~al.}(2011)\citenamefont {Escher},
  \citenamefont {de~Matos~Filho},\ and\ \citenamefont
  {Davidovich}}]{Escher:11}%
  \BibitemOpen
  \bibfield  {author} {\bibinfo {author} {\bibfnamefont {B.~M.}\ \bibnamefont
  {Escher}}, \bibinfo {author} {\bibfnamefont {R.~L.}\ \bibnamefont
  {de~Matos~Filho}}, \ and\ \bibinfo {author} {\bibfnamefont {L.}~\bibnamefont
  {Davidovich}},\ }\href {\doibase 10.1038/nphys1958} {\bibfield  {journal}
  {\bibinfo  {journal} {Nat. Phys.}\ }\textbf {\bibinfo {volume} {7}},\
  \bibinfo {pages} {406} (\bibinfo {year} {2011})}\BibitemShut {NoStop}%
\bibitem [{\citenamefont {Escher}\ \emph {et~al.}(2012)\citenamefont {Escher},
  \citenamefont {Davidovich}, \citenamefont {Zagury},\ and\ \citenamefont
  {de~Matos~Filho}}]{Escher:12}%
  \BibitemOpen
  \bibfield  {author} {\bibinfo {author} {\bibfnamefont {B.~M.}\ \bibnamefont
  {Escher}}, \bibinfo {author} {\bibfnamefont {L.}~\bibnamefont {Davidovich}},
  \bibinfo {author} {\bibfnamefont {N.}~\bibnamefont {Zagury}}, \ and\ \bibinfo
  {author} {\bibfnamefont {R.~L.}\ \bibnamefont {de~Matos~Filho}},\ }\href
  {\doibase 10.1103/PhysRevLett.109.190404} {\bibfield  {journal} {\bibinfo
  {journal} {Phys. Rev. Lett.}\ }\textbf {\bibinfo {volume} {109}},\ \bibinfo
  {pages} {190404} (\bibinfo {year} {2012})}\BibitemShut {NoStop}%
\bibitem [{\citenamefont {Demkowicz-Dobrza{\'n}ski}\ \emph
  {et~al.}(2012)\citenamefont {Demkowicz-Dobrza{\'n}ski}, \citenamefont
  {Ko{\l}ody{\'n}ski},\ and\ \citenamefont {Gu{\c{t}}{\u{a}}}}]{Kolodynski:12}%
  \BibitemOpen
  \bibfield  {author} {\bibinfo {author} {\bibfnamefont {R.}~\bibnamefont
  {Demkowicz-Dobrza{\'n}ski}}, \bibinfo {author} {\bibfnamefont
  {J.}~\bibnamefont {Ko{\l}ody{\'n}ski}}, \ and\ \bibinfo {author}
  {\bibfnamefont {M.}~\bibnamefont {Gu{\c{t}}{\u{a}}}},\ }\href {\doibase
  10.1038/ncomms2067} {\bibfield  {journal} {\bibinfo  {journal} {Nat.
  Commun.}\ }\textbf {\bibinfo {volume} {3}},\ \bibinfo {pages} {1063}
  (\bibinfo {year} {2012})}\BibitemShut {NoStop}%
\bibitem [{\citenamefont {Sekatski}\ \emph
  {et~al.}(2017{\natexlab{a}})\citenamefont {Sekatski}, \citenamefont
  {Skotiniotis}, \citenamefont {Ko{\l{}}ody{\'{n}}ski},\ and\ \citenamefont
  {D{\"{u}}r}}]{Sekatski2017quantummetrology}%
  \BibitemOpen
  \bibfield  {author} {\bibinfo {author} {\bibfnamefont {P.}~\bibnamefont
  {Sekatski}}, \bibinfo {author} {\bibfnamefont {M.}~\bibnamefont
  {Skotiniotis}}, \bibinfo {author} {\bibfnamefont {J.}~\bibnamefont
  {Ko{\l{}}ody{\'{n}}ski}}, \ and\ \bibinfo {author} {\bibfnamefont
  {W.}~\bibnamefont {D{\"{u}}r}},\ }\href {\doibase 10.22331/q-2017-09-06-27}
  {\bibfield  {journal} {\bibinfo  {journal} {{Quantum}}\ }\textbf {\bibinfo
  {volume} {1}},\ \bibinfo {pages} {27} (\bibinfo {year}
  {2017}{\natexlab{a}})}\BibitemShut {NoStop}%
\bibitem [{\citenamefont {Demkowicz-Dobrza\ifmmode~\acute{n}\else
  \'{n}\fi{}ski}\ \emph {et~al.}(2017)\citenamefont
  {Demkowicz-Dobrza\ifmmode~\acute{n}\else \'{n}\fi{}ski}, \citenamefont
  {Czajkowski},\ and\ \citenamefont {Sekatski}}]{Sekatski2017PRX}%
  \BibitemOpen
  \bibfield  {author} {\bibinfo {author} {\bibfnamefont {R.}~\bibnamefont
  {Demkowicz-Dobrza\ifmmode~\acute{n}\else \'{n}\fi{}ski}}, \bibinfo {author}
  {\bibfnamefont {J.}~\bibnamefont {Czajkowski}}, \ and\ \bibinfo {author}
  {\bibfnamefont {P.}~\bibnamefont {Sekatski}},\ }\href {\doibase
  10.1103/PhysRevX.7.041009} {\bibfield  {journal} {\bibinfo  {journal} {Phys.
  Rev. X}\ }\textbf {\bibinfo {volume} {7}},\ \bibinfo {pages} {041009}
  (\bibinfo {year} {2017})}\BibitemShut {NoStop}%
\bibitem [{\citenamefont {D\"ur}\ \emph {et~al.}(2014)\citenamefont {D\"ur},
  \citenamefont {Skotiniotis}, \citenamefont {Fr\"owis},\ and\ \citenamefont
  {Kraus}}]{Dur_2014}%
  \BibitemOpen
  \bibfield  {author} {\bibinfo {author} {\bibfnamefont {W.}~\bibnamefont
  {D\"ur}}, \bibinfo {author} {\bibfnamefont {M.}~\bibnamefont {Skotiniotis}},
  \bibinfo {author} {\bibfnamefont {F.}~\bibnamefont {Fr\"owis}}, \ and\
  \bibinfo {author} {\bibfnamefont {B.}~\bibnamefont {Kraus}},\ }\href
  {\doibase 10.1103/PhysRevLett.112.080801} {\bibfield  {journal} {\bibinfo
  {journal} {Phys. Rev. Lett.}\ }\textbf {\bibinfo {volume} {112}},\ \bibinfo
  {pages} {080801} (\bibinfo {year} {2014})}\BibitemShut {NoStop}%
\bibitem [{\citenamefont {Arrad}\ \emph {et~al.}(2014)\citenamefont {Arrad},
  \citenamefont {Vinkler}, \citenamefont {Aharonov},\ and\ \citenamefont
  {Retzker}}]{Arrad_2014}%
  \BibitemOpen
  \bibfield  {author} {\bibinfo {author} {\bibfnamefont {G.}~\bibnamefont
  {Arrad}}, \bibinfo {author} {\bibfnamefont {Y.}~\bibnamefont {Vinkler}},
  \bibinfo {author} {\bibfnamefont {D.}~\bibnamefont {Aharonov}}, \ and\
  \bibinfo {author} {\bibfnamefont {A.}~\bibnamefont {Retzker}},\ }\href
  {\doibase 10.1103/PhysRevLett.112.150801} {\bibfield  {journal} {\bibinfo
  {journal} {Phys. Rev. Lett.}\ }\textbf {\bibinfo {volume} {112}},\ \bibinfo
  {pages} {150801} (\bibinfo {year} {2014})}\BibitemShut {NoStop}%
\bibitem [{\citenamefont {Kessler}\ \emph {et~al.}(2014)\citenamefont
  {Kessler}, \citenamefont {Lovchinsky}, \citenamefont {Sushkov},\ and\
  \citenamefont {Lukin}}]{Kessler_2014}%
  \BibitemOpen
  \bibfield  {author} {\bibinfo {author} {\bibfnamefont {E.~M.}\ \bibnamefont
  {Kessler}}, \bibinfo {author} {\bibfnamefont {I.}~\bibnamefont {Lovchinsky}},
  \bibinfo {author} {\bibfnamefont {A.~O.}\ \bibnamefont {Sushkov}}, \ and\
  \bibinfo {author} {\bibfnamefont {M.~D.}\ \bibnamefont {Lukin}},\ }\href
  {\doibase 10.1103/PhysRevLett.112.150802} {\bibfield  {journal} {\bibinfo
  {journal} {Phys. Rev. Lett.}\ }\textbf {\bibinfo {volume} {112}},\ \bibinfo
  {pages} {150802} (\bibinfo {year} {2014})}\BibitemShut {NoStop}%
\bibitem [{\citenamefont {Sekatski}\ \emph {et~al.}(2016)\citenamefont
  {Sekatski}, \citenamefont {Skotiniotis},\ and\ \citenamefont
  {D\"ur}}]{Sekatski_2016}%
  \BibitemOpen
  \bibfield  {author} {\bibinfo {author} {\bibfnamefont {P.}~\bibnamefont
  {Sekatski}}, \bibinfo {author} {\bibfnamefont {M.}~\bibnamefont
  {Skotiniotis}}, \ and\ \bibinfo {author} {\bibfnamefont {W.}~\bibnamefont
  {D\"ur}},\ }\href {\doibase 10.1088/1367-2630/18/7/073034} {\bibfield
  {journal} {\bibinfo  {journal} {New Journal of Physics}\ }\textbf {\bibinfo
  {volume} {18}},\ \bibinfo {pages} {073034} (\bibinfo {year}
  {2016})}\BibitemShut {NoStop}%
\bibitem [{\citenamefont {Zhou}\ \emph {et~al.}(2018)\citenamefont {Zhou},
  \citenamefont {Zhang}, \citenamefont {Preskill},\ and\ \citenamefont
  {Jiang}}]{Zhou2018}%
  \BibitemOpen
  \bibfield  {author} {\bibinfo {author} {\bibfnamefont {S.}~\bibnamefont
  {Zhou}}, \bibinfo {author} {\bibfnamefont {M.}~\bibnamefont {Zhang}},
  \bibinfo {author} {\bibfnamefont {J.}~\bibnamefont {Preskill}}, \ and\
  \bibinfo {author} {\bibfnamefont {L.}~\bibnamefont {Jiang}},\ }\href
  {\doibase 10.1038/s41467-017-02510-3} {\bibfield  {journal} {\bibinfo
  {journal} {Nature Communications}\ }\textbf {\bibinfo {volume} {9}},\
  \bibinfo {pages} {78} (\bibinfo {year} {2018})}\BibitemShut {NoStop}%
\bibitem [{\citenamefont {Sekatski}\ \emph
  {et~al.}(2017{\natexlab{b}})\citenamefont {Sekatski}, \citenamefont
  {Skotiniotis},\ and\ \citenamefont {D\"ur}}]{Sekatski2017singlequbit}%
  \BibitemOpen
  \bibfield  {author} {\bibinfo {author} {\bibfnamefont {P.}~\bibnamefont
  {Sekatski}}, \bibinfo {author} {\bibfnamefont {M.}~\bibnamefont
  {Skotiniotis}}, \ and\ \bibinfo {author} {\bibfnamefont {W.}~\bibnamefont
  {D\"ur}},\ }\href {\doibase 10.1103/PhysRevLett.118.170801} {\bibfield
  {journal} {\bibinfo  {journal} {Phys. Rev. Lett.}\ }\textbf {\bibinfo
  {volume} {118}},\ \bibinfo {pages} {170801} (\bibinfo {year}
  {2017}{\natexlab{b}})}\BibitemShut {NoStop}%
\bibitem [{\citenamefont {Holevo}(1973)}]{Holevo:73}%
  \BibitemOpen
  \bibfield  {author} {\bibinfo {author} {\bibfnamefont {A.~S.}\ \bibnamefont
  {Holevo}},\ }\href@noop {} {\bibfield  {journal} {\bibinfo  {journal}
  {Problemy Peredachi Informatsii}\ }\textbf {\bibinfo {volume} {9}},\ \bibinfo
  {pages} {3} (\bibinfo {year} {1973})}\BibitemShut {NoStop}%
\bibitem [{\citenamefont {Berry}\ and\ \citenamefont
  {Wiseman}(2000)}]{Berry2000}%
  \BibitemOpen
  \bibfield  {author} {\bibinfo {author} {\bibfnamefont {D.~W.}\ \bibnamefont
  {Berry}}\ and\ \bibinfo {author} {\bibfnamefont {H.~M.}\ \bibnamefont
  {Wiseman}},\ }\href {\doibase 10.1103/PhysRevLett.85.5098} {\bibfield
  {journal} {\bibinfo  {journal} {Phys. Rev. Lett.}\ }\textbf {\bibinfo
  {volume} {85}},\ \bibinfo {pages} {5098} (\bibinfo {year}
  {2000})}\BibitemShut {NoStop}%
\bibitem [{\citenamefont {W\"olk}\ \emph {et~al.}()\citenamefont {W\"olk},
  \citenamefont {Sekatski},\ and\ \citenamefont
  {D\"ur}}]{WoelkSekatski_inpreparation}%
  \BibitemOpen
  \bibfield  {author} {\bibinfo {author} {\bibfnamefont {S.}~\bibnamefont
  {W\"olk}}, \bibinfo {author} {\bibfnamefont {P.}~\bibnamefont {Sekatski}}, \
  and\ \bibinfo {author} {\bibfnamefont {W.}~\bibnamefont {D\"ur}},\
  }\href@noop {} {}\bibinfo {note} {\textit{in preparation}}\BibitemShut
  {NoStop}%
\bibitem [{\citenamefont {Bartlett}\ \emph {et~al.}(2007)\citenamefont
  {Bartlett}, \citenamefont {Rudolph},\ and\ \citenamefont
  {Spekkens}}]{BartlettRMP}%
  \BibitemOpen
  \bibfield  {author} {\bibinfo {author} {\bibfnamefont {S.~D.}\ \bibnamefont
  {Bartlett}}, \bibinfo {author} {\bibfnamefont {T.}~\bibnamefont {Rudolph}}, \
  and\ \bibinfo {author} {\bibfnamefont {R.~W.}\ \bibnamefont {Spekkens}},\
  }\href {\doibase 10.1103/RevModPhys.79.555} {\bibfield  {journal} {\bibinfo
  {journal} {Rev. Mod. Phys.}\ }\textbf {\bibinfo {volume} {79}},\ \bibinfo
  {pages} {555} (\bibinfo {year} {2007})}\BibitemShut {NoStop}%
\end{thebibliography}%

\begin{appendix}	
\section{Optimal QFI for the signal staying insensitive to noise}

Here we show that if the generator of the signal $G_{k_*}$ is linearly independent from the noise, i.e. the set of generations $\{G\}_{k\neq k_*}$, the superposition state of Eq.~\eqref{eq:probe_state} with ${\bf s}_*$ given by ${\bf s}_* = \textrm{argmax}_{\bf s}\{{\bf f}_{k_*}^T {\bf s} |\,{\bf f}_{k}^T {\bf s} = 0 \quad \forall \quad k\neq k_*\}$
is insensitive to noise and gives the maximal QFI with respect to the signal among all states insensitive to noise.

Following the discussion in the main text, define a vector space $V_\textrm{noise} = \textrm{span}\{ {\bf f}_k\}_{k\neq k_*}$, and assume that ${\bf f}_{k_*}$ is linearly independent from the set of all other vectors ${\bf f}_k$. Then it can be uniquely decomposed as ${\bf f}_{k_*} = {\bf f}_\perp + {\bf f}_\parallel$ with ${\bf f}_\parallel\in V_\textrm{noise}$ and ${\bf f}_\perp^T {\bf f}_k = 0$ for $k\neq k_*$. In the same way the vectors can be decomposed in orthogonal components ${\bf s}= s_* {\bf f}_\perp + {\bf s}_\parallel +{\bf s}_\textrm{ext}$, where ${\bf s}_\parallel \in V_\text{noise}$ and ${\bf s}_\textrm{ext}$ orthogonal to both ${\bf f}_\perp$ and $V_\textrm{noise}$. The insensitivity to noise constraints can be satisfied by chosing ${\bf s}_\parallel = {\bf r}_\parallel = {\bf v}_\parallel$. To obtain the optimal strategy it remains to find the two vectors 
\begin{align}
{\bf s}= s_* {\bf f}_\perp + {\bf v}_\parallel +{\bf s}_\textrm{ext} \in O_{\bf n}\\ {\bf r}= r_* {\bf f}_\perp + {\bf v}_\parallel +{\bf r}_\textrm{ext}\in O_{\bf n}
\end{align}
maximizing the QFI $=(s_* - r_*)^2$. To do so, we first get rid of the components ${\bf s}_\textrm{ext}$ and ${\bf r}_\textrm{ext}$ by projecting the orthotope $O_{\bf n}$ onto the subspace spanned by ${\bf f}_k$ and all the vectors in $V_\textrm{noise}$. This gives a polytope $P_{\bf n}$ that is symmetric under inversion, i.e. for any vector ${\bf v} \in P_{\bf n}$ the opposite vector $-{\bf v}$ is also in $P_{\bf n}$. The optimization becomes 
\be
\max \{ (s_* -r*)^2 \Big| {\bf s}= s_* {\bf f}_\perp + {\bf v}_\parallel \in P_{\bf n}, {\bf r}= r_* {\bf f}_\perp + {\bf v}_\parallel \in P_{\bf n}\}.
\ee
Because of the symmetry of $P_{\bf n}$ the maximal difference is always attained for ${\bf v}_\parallel =0$. This is because for any two such two vectors ${\bf s}$ and ${\bf r}$ in $P_n$ their inverses $-{\bf s}$ and $-{\bf r}$ are also inside the polytope. By convexity, so are the vectors 
\begin{align}
    {\bf s}'&= \frac{1}{2} ({\bf s}-{\bf r} ) =\frac{1}{2} (s_* -r_*){\bf f}_\perp = s_*' {\bf f}_\perp\\
     {\bf r}'&= \frac{1}{2} ({\bf r}-{\bf s} ) = -\frac{1}{2} (s_* -r_*){\bf f}_\perp = r_*' {\bf f}_\perp,
\end{align}
that yield the quantum Fisher information  $(s_*'-r_*')^2 = (s_*-r_*)^2$ and satisfy ${\bf v}_\parallel=0$. Furthermore, we get $r_* = - s_*$. By the same inversion symmetry it follows that the state for the state attaining the value 
${\bf s} = s_*'{\bf f_k} + {\bf s}_\textrm{ext} \in O_{\bf n}$
attaining the value $s_*$, it's inverse attains $-s_*'$. Hence, the optimal vectors can be taken to be
\begin{align}
{\bf s} &= s_* {\bf f_\perp} + {\bf s}_\textrm{ext}  \\
{\bf r} &= -{\bf s}.
\end{align}


\section{Action of noise as orthogonal projection on decoherence free subspaces.}

In the noiseless case the evolution of the probes for a time $t$ is described by a global unitary operator $U= \exp\left(i t\sum_k \alpha_k G_k\right) = \exp\left(i\sum_k \Phi_k G_k\right)$. 
Suppose now, that one is only interested in measuring the signal $\alpha_*$, while the other signals $\alpha_k$ fluctuate and therefore add noise to the final state, and we refer to those as noise processes. In particular, we consider the most pessimistic case where for any finite evolution time $t$ the values $\alpha_k$ (for $k\neq k_*$) fluctuate in such a way that the parameters $\Phi_k$ are independent random variables sampled anew for each experimental run. In other words, at each run the effect of each noise process on the state of the probes is to apply a unitary $e^{i \Phi_{k} G_{k}}$ with a random value $\Phi_{k}$ unknown to the experimentor. Hence, the evolution is no longer unitary, but is given by a completely positive trace preserving map 
\be
\cE_{\Phi_{k_*}}[\bullet] = e^{i \Phi_{k_*} G_{k_*}} \left(\underbrace{\mathcal{T}_{G_1}\circ\dots\circ \mathcal{T}_{G_J}}_{k\neq k_*}[\bullet]\right) e^{-i \Phi_{k_*} G_{k_*}},
\ee
where each $\mathcal{T}_{G_k}[\bullet]=\int d\Phi_k\, e^{i \Phi_{k} G_{k}}\bullet e^{-i \Phi_{k} G_{k}}$, with $\int d\Phi_k=1$, is a twirling map \cite{BartlettRMP}. Such a twirling map destroys the coherence between any two eigenstates of $G_k$ that have different eigenvalues. Hence, it can equivalently represented as an orthogonal projection onto blocks that only contains product states that are degenerate for all the noise generators \footnote{We reming the reader, that all the generators are diagonal in the basis of product states in computation basis.}
\begin{align}
    {\bf T}[\bullet] = \underbrace{\mathcal{T}_{G_1}\circ\dots\circ \mathcal{T}_{G_J}}_{k\neq k_*}[\bullet] =\bigoplus_{\bm \lambda} \Pi_{\bm \lambda}\bullet \Pi_{\bm \lambda},
\end{align}
where
\be
\Pi_{\bm \lambda} = \sum_{\bf s} \proj{\bf s} \quad \textrm{s.t.} \quad G_k\ket{\bf s} = \lambda_k \ket{\bf s} \quad \forall k\neq k_*
\ee 
and ${\bm \lambda} = (\lambda_1 \dots \lambda_k)$ labels the blocks by the eigenvalues of all noise generators. 

Such a noise process destroys any coherence between two states belonging to different blocks ${\bm \lambda}$ and ${\bm \lambda}'$, that is initially present in the probe state.
Hence, it is not only convenient to prepare the probes in a noise incentive states, but in fact any probe state gets projected into a mixture of noise insensitive states by the evolution.


\section{Examples}

\paragraph{Taylor expansion.---}

The Taylor expansion is create by the generating functions $f_k(r)=(r/r_0)^k$ where $r_0$ defines a length scale. Furthermore, we assume that our sensors are placed at the positions $r_j/r_0 \in \lbrace 0, \pm 1 , \pm 2\rbrace$. In this case, all vectors $\mathbf{f}_k$ with odd $k$ are orthogonal to all vectors with even $k$ and vice versa. However, the subgroup of all odd (even) vectors are not orthogonal within their subgroup. As an example, we want to measure $\Phi_3$ without noise from the order $k \in \lbrace 0,1,2,4\rbrace$. The vector $\mathbf{f}_3=((-2)^3,(-1)^3,0,1^3,2^3)^T$ is already orthogonal to $\mathbf{f}_0$, $\mathbf{f}_2$ and $\mathbf{f}_4$ but not to $\mathbf{f}_1=(-2,-1,0,1,2)^T$. With the help of the Gram-Schmidt process we obtain the orthogonal vector ${\bf f}_\perp=(-1,2,0,-2,1)^T$ which is orthogonal to all $\mathbf{f}_k$ with $k\neq 3$ and has maximal overlap with $\mathbf{f}_3$. By placing $n=\floor{N/6}$ qubits at positions $r_j/r_0=\pm2$, $2n$ qubits at positions $r_j/r_0=\pm1$  and zero qubits at $r_j/r_0=0$ we can create the optimal probe state
\begin{equation}
\ket{\psi_3^\text{T}}=\tfrac{1}{\sqrt 2} \left(\ket{-n,2n,0,-2n,n}+\ket{n,-2n,0,2n,-n}
\right).
\end{equation}
Note that in this case all values are integer and can be matched by placing an appropriate number of qubits at each sensor position. If the number of sensing systems is however restricted, e.g. to a single qubit per sensor, then local control is required to obtain required effective eigenvalues.

\paragraph{Fourier Expansion.---}

In the following, we consider the generating functions given by $f_k(r)=\sin[k\pi (r/r_0)]$ which corresponds to standing waves with boundary conditions $B(r)=0$ at positions $r/r_0=0$ and $r/r_0=1$. Without noise, it is optimal the place the sensors at positions $r_j$ with maximal absolute field strength $|f_{k_\ast}(r_j)|=1$ that is
\begin{equation}
r_j=\frac{\pi}{k_\ast}\left(j-\frac{1}{2}\right)\; \text{with} \; j=1\cdots k_\ast.
\end{equation}
As a result, we obtain a coefficient vector $\mathbf{f}_{k_\ast}=(1,-1,1,-1,\cdots)$ with alternating entries and maximal absolute values. The optimal probe state to measure $\Phi_{k_\ast}$
is then given by
\begin{equation}
\ket{\psi_{k_\ast}^\text{F}}=\tfrac{1}{\sqrt 2}\left(\ket{n,-n, n \cdots} +\ket{-n,n, -n \cdots}
\right),
\end{equation}
if each sensor consists of $n=N/k_\ast$ qubits.

In addition, this state is eigenstate of all generators $\hat{G}_k$ with $k\neq k_\ast(1+2m)$ with $m \in \mathbb{N}$ since
\be
\sum\limits_{j=1}^{k_0} \frac{(-1)^j}{k_0} \sin\left[\pi \frac{k}{k_0}(j-\frac{1}{2})\right] = \left\lbrace \begin{array}{cc} (-1)^m & k=k_0(1+2m) \\ 0 & \text{else}\end{array}\right. .
\ee
As a consequence, this probe state is already insensitive to all noise sources with $k\neq k_0(1+2m)$ and especially $k<k_\ast$. Notice that one can also achieve insensitivity against noise sources with a higher $k$ than the signal, which might require the usage of additional sensor positions.

\paragraph{Point sources.---}

\begin{figure}[h]
    \includegraphics[width=0.7 \columnwidth]{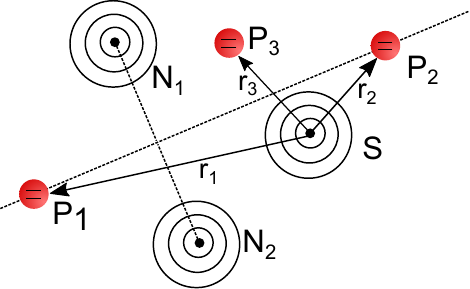}
    \caption{The strength of the signal $S$ should be determined via three sensors $P_j$. The two noise sources $N_1$ and $N_2$ are equidistant from  the sensor $P_1$ as well as $P_2$. }
    \label{Fig_point_sources}
\end{figure}

\begin{figure}[h]
    \includegraphics[width=0.6 \columnwidth]{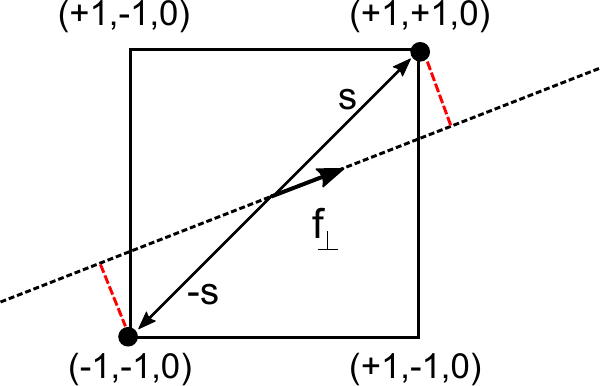}
    \caption{The polytope defined by $(\pm 1, \pm 1, 0)$ describes the decoherence free subspace if there is only a single qubit present at each sensor. The two states ${\bf s}=\ket{1,1,0}$ and $-\bf s$ lead a maximal projection onto $f_\perp$  of states within the polytope.     }
    \label{Fig_polytope}
\end{figure}

We now investigate a case where the noise and the signal field are created by point sources with a $1/r^2$ distance dependence. We consider a specific, simple setting in order to illustrate the construction of optimal states for sensing. We assume that the noise sources $N_1$ and $N_2$ have equal strength but opposite sign. As a consequence, there exist a plane incorporating all points where the noise of $N_1$ and $N_2$ cancel each other. Furthermore, we assume that the two sensors $P_1$ and $P_2$ lie exactly in this plane as depicted in Fig. \ref{Fig_point_sources}. A third sensor $P_3$ as well as the signal source $S$ are positioned somewhere outside this plane. 

The noise sources can be described by a single effective vector ${\bf f}_N={\bf f}_{N_1}-{\bf f}_{N_2}=(0,0,1)^T$. The vector describing the generating function of the signal source is given by ${\bf f}_S=(1/r_1^2,1/r_2^2,1/r_3^2)^T$ with the orthogonal component ${\bf f}_\perp=(r_2^2,r_1^2,0)^T$. However, the decoherence free subspace is two dimensional and is spanned by ${\bf f}_\perp$ and ${\bf s}_\text{ext}=(r_1^2,-r_2^2,0)^T$. In this case, the optimal probe state is not described by ${\bf s}\parallel {\bf f}_\perp$ if the maximal number of spins at the sensors $P_{1/2}$ are limited by $n_{1/2}$ with $n_1/n_2\neq r_2^2/r_1^2$. To construct the optimal ${\bf s}$ we need to project the polytope $O_n=[-n_1,n_1]\times [-n_2,n_2]\times[-n_3,n_3] $ onto the decoherence free subspace. The polytope $O_n$ is given by a cube for $n_1=n_2=n_3$ and its projection on the decoherence free subspace is determined by a square with the extremal points $(\pm 1, \pm 1, 0)$ for $n_1=n_2=1$ as depicted in Fig.\ref{Fig_polytope}. Here, we have chosen $s_3=0$ because the optimal state can be always obtained by choosing $v_\parallel=0$ (see Appendix 1) leading to $s_3=0$ in our case. The maximal projection $\bf{s}_\ast={\bf f}_S^T {\bf s}$ is obtained for the two extremal points of the polytope described by $\pm{\bf s}=(1,1,0)^T$ as shown in Fig. \ref{Fig_polytope}.

Let us briefly come back to the general case of point emitters described by field configurations
\be
f_k({\bf r}) = \frac{1}{|{\bf r}- {\bf R}_k|^\beta},
\ee
 where all such functions are linearly independent as long as the locations of all the sources are different ${\bf R}_k \neq {\bf R}_{k'}$. In a situation where the size of the sensors network is small as compared to the distance to the nearby emitter, that is $|{\bf r}_j - {\bf r}_\ell|\ll |{\bf R}_k|$, it is natural to expand the signal configurations in a Taylor series around a point ${\bf r}_0$
\be
f_k({\bf r})= f_k^{(0)} +  ({\bf r}-{\bf r}_0)^T {\bf f}_k^{(1)}+\dots.
\ee
However, in contrast to the original function, the expansions up to a fixed order $n$ are not linearly independent in general: there are $\frac{(n+1)(n+2)}{2}$ and  $\frac{(n+1)(n+2)(n+3)}{6}$ linearly independent polynomials of degree $n$ for ${\bf r}\in \mathbb{R}^2$ (2 variables) and ${\bf r}\in \mathbb{R}^3$ (3 variables) respectively. This shows that $n$ has to be taken large enough if the analysis is done from the Taylor expansion perspective. In addition, one sees that in such a situation the signal is rather weak after projection in the noise-insensitive subspace: the difference between the sources will only appear in the last orders of the expansion.

\end{appendix}



\end{document}